\documentclass[aps,pre,reprint,nofootinbib,superscriptaddress]{revtex4-2}

\usepackage{amsmath,amssymb,bm}
\usepackage{mathtools}
\usepackage{booktabs}
\usepackage{graphicx}
\graphicspath{{outputs/figures/}}
\usepackage[colorlinks=true,linkcolor=blue,citecolor=blue,urlcolor=blue]{hyperref}

\newcommand{\dd}{\mathrm d}
\newcommand{\Dt}{\Delta t}
\newcommand{\tr}{\operatorname{tr}}
\newcommand{\Jsig}{\mathcal J_\sigma}
\newcommand{\Msig}{M_\sigma}
\newcommand{\eff}{\mathrm{eff}}
\newcommand{\disc}{\mathrm{disc}}

\newcommand{\full}{\mathrm{full}}
\newcommand{\diag}{\mathrm{diag}}

\begin{document}

\title{Matrix-noise Jacobians in stochastic-calculus inference and optimal paths}

\author{Surachate Limkumnerd}
\affiliation{Department of Physics, Faculty of Science, Chulalongkorn University, Bangkok 10330, Thailand}

\date{\today}

\begin{abstract}
Multiplicative noise makes stochastic dynamics depend on how the white-noise limit is interpreted.  In multidimensional systems with matrix-valued noise amplitudes \(\sigma(x)\), this dependence includes a local Jacobian contribution that is absent from the scalar examples most often used to build intuition.  We formulate a finite-step path-likelihood framework for \(\theta\)-discretized diffusions and show that its short-time expansion isolates the scalar \(\Jsig=\partial_j\sigma_{ik}\partial_i\sigma_{jk}-(\partial_i\sigma_{ik})(\partial_l\sigma_{lk})\).  For a specified noise-amplitude representation \(\sigma\), this quantity vanishes in one-dimensional, scalar-isotropic, and strictly diagonal cases, but can survive when state-dependent noise directions mix different components.  We then test its consequences using paired comparisons that hold the drift, diffusion matrix, interpolation point, and Gaussian increment term fixed.  In Model A, removing only the off-diagonal determinant contribution produces a shift in the fitted stochastic prescription that vanishes when \(\Jsig=0\).  In Model B, removing the corresponding state-dependent action term changes a stable optimized transition path.  These results show that a genuinely matrix-noise part of the short-time path measure can survive the scalar cancellations familiar from simpler settings and produce measurable changes in fitted stochastic prescriptions and Onsager--Machlup paths.
\end{abstract}

\maketitle

\section{Introduction}

Stochastic differential equations with multiplicative noise are not fully specified by writing a drift and a noise amplitude.  One must also specify how the white-noise limit is interpreted.  The It\^o, Stratonovich, and H\"anggi--Klimontovich prescriptions are often introduced as alternative calculi, but in coarse-grained physical models the prescription can encode information about unresolved microscopic dynamics.  Adiabatic elimination with state-dependent damping, simultaneous small-mass and colored-noise limits, and thermodynamically consistent descriptions of state-dependent diffusion all show that the effective drift depends on how fast variables, memory, and noise correlations are removed \cite{SanchoSanMiguelDurr1982,KupfermanPavliotisStuart2004,LauLubensky2007,VolpeWehr2016}.  In this sense the stochastic prescription is not merely notation; it is a parameter summarizing hidden time scales and experimental resolution.

This point has also become experimentally and inferentially important.  Pesce \emph{et al.} demonstrated that a noisy electric circuit can shift from Stratonovich-like to It\^o-like behavior as operating conditions change, specifically as the ratio of noise correlation time to feedback delay time varies \cite{Pesce2013}.  More recent work on heterogeneous diffusion asks how the interpretation should be chosen from single-particle trajectories, especially when diffusivity varies in space across phase-separated environments \cite{PachecoPozo2024,Liu2026}.  Related force-inference problems in heterogeneous environments face the same structural difficulty: a spurious force arising from diffusivity gradients can be entangled with the true force in measured trajectories \cite{Serov2020}.  These studies motivate treating the stochastic prescription as an inferential object, but they also expose a limitation of density-level information.  The Fokker--Planck generator identifies only a convention-dependent combination of drift and noise-induced drift.  Without additional path-level or physical information, the prescription cannot be separated from the drift.

Path-integral and Onsager--Machlup formulations provide a natural language for that additional information.  The classical Onsager--Machlup theory and its later extensions relate diffusion processes to path weights and most probable paths \cite{OnsagerMachlup1953,Graham1977,DurrBach1978}.  For multiplicative noise, however, the construction is delicate because discretization, stochastic chain rules, Jacobians, and changes of variables are tied together.  Symmetric path-integral treatments of multiplicative noise have clarified how prescription-dependent terms enter the action \cite{Arnold2000}.  Supersymmetric and vector-variable functional formulations have further developed this picture \cite{ArenasBarci2010,ArenasBarci2012,MorenoArenasBarci2015}.  Cugliandolo and Lecomte showed explicitly that the usual rules of calculus cannot be transferred naively into the Onsager--Machlup path integral and developed a consistent path-integral calculus in one degree of freedom \cite{CugliandoloLecomte2017}.  Later covariant and discretized approaches extended this viewpoint and emphasized the geometric and discretization structure of stochastic path integrals \cite{CugliandoloLecomteVanWijland2019,DePireyCugliandoloLecomteVanWijland2022}.  Moreno and Barci, and related work, addressed transition probabilities for multiplicative-noise processes and the role of local Jacobian factors in finite-time propagators \cite{MorenoBarci2019,AbrilBermudez2025}.

The central claim of this work is that, once the noise amplitude is genuinely matrix-valued, the short-time path measure contains a Jacobian remainder that is absent from scalar intuition but can nevertheless be isolated and tested operationally.  For a specified noise-channel amplitude \(\sigma_{ik}(x)\), rather than only its covariance \(K=\sigma\sigma^T\), the relevant scalar remainder is
\begin{equation}
\Jsig=
\partial_j\sigma_{ik}\partial_i\sigma_{jk}
-
(\partial_i\sigma_{ik})(\partial_l\sigma_{lk}).
\label{eq:Jdef}
\end{equation}
Thus \(\Jsig\) is a diagnostic of the chosen noise-amplitude representation fixed by the stochastic model or microscopic regularization; it is not claimed to be a coordinate-free scalar determined by \(K\) alone.  It separates the genuinely matrix-valued short-time Jacobian contribution from pieces already visible in scalar intuition.  This term vanishes in one-dimensional, scalar-isotropic, and strictly diagonal examples, but it survives when the local noise directions mix state components.

Such structures naturally arise in statistical and nonequilibrium settings where the noise is tied to an anisotropic diffusion tensor, a mobility tensor, or state-dependent correlated fluctuations.  In curvilinear or non-principal coordinates, in coarse-grained hydrodynamic variables, after eliminating fast variables, or after projecting stochastic dynamics onto constrained manifolds, one expects genuinely matrix-valued noise amplitudes rather than a globally scalar noise strength.  Nonequilibrium coarse-graining can similarly produce state-dependent noise directions whose covariance and amplitude cannot be diagonalized by a single global coordinate choice.  In these cases off-diagonal derivative structures can survive, so the scalar \(\Jsig\) need not vanish.  The present work does not attempt a full classification of such systems; instead it isolates one structurally identifiable remainder and studies its controlled consequences in minimal examples.

This leads to three concrete steps.  We write the finite-step likelihood for a \(\theta\)-discretized multidimensional diffusion and show how its short-time expansion contains \(\Jsig\).  We identify common noise classes for which \(\Jsig\) vanishes.  We then use paired controls in which the drift, diffusion matrix, interpolation point, and Gaussian increment term are held fixed: in Model A a reduced-determinant control shifts the fitted prescription, and in Model B the corresponding reduced action changes a stable optimized transition path.  Together, these examples show that the matrix-Jacobian contribution can affect fitted prescriptions and optimized paths without claiming an exhaustive classification of multidimensional multiplicative-noise systems.

The paper is organized as follows.  Section~\ref{sec:generator} shows that the Fokker--Planck generator cannot identify the stochastic prescription separately from the drift.  Section~\ref{sec:action} derives the short-time action and the finite-step discrete kernel used for inference.  Section~\ref{sec:visibility} identifies common cases in which the matrix-Jacobian scalar is hidden or visible.  Section~\ref{sec:inference} defines path-likelihood inference of \(\theta\) and presents the first numerical diagnostic.  Section~\ref{sec:paths} derives the Hamiltonian structure of the optimal-path problem and presents the second diagnostic.  Section~\ref{sec:discussion} discusses the scope and limitations of the result.  Appendices~\ref{sec:S-action}--\ref{sec:S-modelB} give the determinant expansion, short-time action, and numerical implementation details.

\section{Generator equivalence}
\label{sec:generator}

Consider the \(d\)-dimensional stochastic differential equation \(\dd x_i = F_i(x)\,\dd t + \sigma_{ik}(x)\,\dd W_k\), interpreted using a \(\theta\)-prescription.  Throughout the main text, repeated Latin indices are summed: \(i,j,l,m,p\) label state components and \(k,n\) label noise channels.  We write \(\partial_i=\partial/\partial x_i\), \(K_{ij}=\sigma_{ik}\sigma_{jk}\), and \(K^{-1}_{ij}\) for the inverse diffusion matrix.  When \(\sigma\) is invertible, \(\sigma^{-1}_{ki}\) denotes the inverse matrix, with \(\sigma^{-1}_{ki}\sigma_{il}=\delta_{kl}\) and \(\sigma_{ik}\sigma^{-1}_{kj}=\delta_{ij}\).  A time-discrete form is
\begin{equation}
x_i^{n+1}-x_i^n =
F_i(\bar x_n)\Dt+
\sigma_{ik}(\bar x_n)\Delta W_k^n ,
\label{eq:theta-update}
\end{equation}
with \(\bar x_n=(1-\theta)x_n+\theta x_{n+1}\).
Here \(\theta=0\) corresponds to It\^o, \(\theta=1/2\) to Stratonovich, and \(\theta=1\) to the anti-It\^o or H\"anggi--Klimontovich convention.

Expanding Eq.~\eqref{eq:theta-update} to It\^o form gives \(\dd x_i = A_i(x)\,\dd t+\sigma_{ik}(x)\,\dd W_k^{\mathrm{Ito}}\), where
\begin{equation*}
A_i(x)=F_i(x)+\theta B_i(x),
\qquad
B_i(x)=\sigma_{jk}(x)\partial_j\sigma_{ik}(x).
\end{equation*}
Thus the Fokker--Planck equation is
\begin{equation}
\partial_t\rho
=
-\partial_i[(F_i+\theta B_i)\rho]
+
\frac12\partial_i\partial_j[K_{ij}\rho],
\label{eq:fpe}
\end{equation}
where \(K_{ij}=\sigma_{ik}\sigma_{jk}\).

Equation~\eqref{eq:fpe} depends on \(F_i\) and \(\theta\) only through the combination \(A_i=F_i+\theta B_i\).  Therefore two descriptions \((F^{(\theta)},\theta)\) and \((F^{(\theta')},\theta')\) generate the same Fokker--Planck operator if
\begin{equation*}
F_i^{(\theta')}(x)
=
F_i^{(\theta)}(x)+(\theta-\theta')B_i(x).
\end{equation*}
This elementary equivalence is central within the Markovian diffusion class with smooth coefficients: a density evolution or steady state alone cannot identify the stochastic prescription separately from the drift unless additional physical information constrains the model.  Path-level likelihoods provide one such additional structure.

\section{Short-time action and discrete kernel}
\label{sec:action}

It is useful to write Eq.~\eqref{eq:theta-update} with white-noise variables \(\eta_k=\Delta W_k/\Dt\):
\begin{equation}
x_i^+ = x_i + F_i(\bar x)\Dt+\sigma_{ik}(\bar x)\eta_k\Dt ,
\qquad
\bar x=(1-\theta)x+\theta x^+ .
\label{eq:white-update}
\end{equation}
The Gaussian noise density is
\begin{equation*}
P(\eta)
=
\left(\frac{\Dt}{2\pi}\right)^{d/2}
\exp\left[-\frac{\Dt}{2}\eta_k\eta_k\right].
\end{equation*}
Let \(J_{il}=\partial x_i^+/\partial \eta_l\).  This finite-dimensional Jacobian matrix \(J\) should not be confused with the scalar \(\Jsig\) defined in Eq.~\eqref{eq:Jdef}.  Differentiating Eq.~\eqref{eq:white-update} gives, up to transpose conventions that do not affect the determinant, \(J=(I-C)^{-1}\sigma\Dt\), where \(C_{ij}=\theta\Dt[\partial_jF_i+\partial_j\sigma_{ik}\eta_k]\).  Thus \(1/|\det J|=\det(I-C)/[|\det\sigma|(\Dt)^d]\).

Because \(\eta_k\sim \Dt^{-1/2}\), the multiplicative-noise part of \(C\) is \(O(\Dt^{1/2})\), and the term \(\tr C^2\) contributes at order \(\Dt\).  Using
\(\log\det(I-C)=-\tr C-\frac12\tr C^2+O(\Dt^{3/2})\),
one obtains
\begin{multline*}
\log\det(I-C)
=
-\theta\partial_iF_i\Dt
-\theta a_k\eta_k\Dt \\
-\frac{\theta^2}{2}\Msig\Dt
+
O(\Dt^{3/2}),
\end{multline*}
where \(a_k=\partial_i\sigma_{ik}\) and \(\Msig=\partial_j\sigma_{ik}\partial_i\sigma_{jk}\).

With \(v_i=(x_i^+-x_i)/\Dt\) and \(\eta_k=\sigma^{-1}_{ki}(v_i-F_i)\), the short-time kernel can be written as \(P_\theta(x^+|x)\simeq\mathcal N(\Dt,\bar x)\exp[-\Dt\,\mathcal L_\theta(\bar x,v)]\), where \(\mathcal N(\Dt,\bar x)\) collects the Gaussian normalization and the local factor \(|\det\sigma(\bar x)|^{-1}\).
Completing the square gives
\begin{widetext}
\begin{equation}
\mathcal L_\theta(x,v)
=
\frac12
\left(
v_i-F_i+\theta\sigma_{ik}\partial_m\sigma_{mk}
\right)
\sigma^{-1}_{li}\sigma^{-1}_{lj}
\left(
v_j-F_j+\theta\sigma_{jn}\partial_p\sigma_{pn}
\right)
+
\theta\,\partial_iF_i
+
\frac{\theta^2}{2}
\left[
\partial_j\sigma_{ik}\partial_i\sigma_{jk}
-
(\partial_i\sigma_{ik})(\partial_l\sigma_{lk})
\right].
\label{eq:Ltheta}
\end{equation}
\end{widetext}
Equation~\eqref{eq:Ltheta} is a local short-time Onsager--Machlup density obtained in the small-\(\Dt\) regime for typical increments \(\Delta x=O(\sqrt{\Dt})\), smooth coefficients, and nonsingular \(\sigma\).  The L\'evy contraction used in deriving it is an asymptotic step, not an exact finite-\(\Dt\) identity.  Therefore Eq.~\eqref{eq:Ltheta} is used below for analytic interpretation and for the continuum variational path calculation, not as the finite-step likelihood used in inference.  The numerical inference in Sec.~\ref{sec:inference} uses the discrete kernel Eq.~\eqref{eq:exact-discrete-kernel}.

For finite-step inference we use the one-step kernel of the discrete scheme before this contraction.  For a candidate \(\theta\), define
\begin{align*}
\bar x_\theta &= x+\theta(x^+-x), \\
\eta_{\theta k}
&=
\sigma^{-1}_{ki}(\bar x_\theta)
\left[
\frac{x_i^+-x_i}{\Dt}
-
F_i(\bar x_\theta)
\right],
\\
C_{\theta,ij}
&=
\theta\Dt[
\partial_jF_i(\bar x_\theta)
+
\partial_j\sigma_{ik}(\bar x_\theta)\eta_{\theta k}
].
\end{align*}
Here \(\eta_{\theta k}\) denotes the \(k\)-th noise-channel component of the vector \(\eta_\theta\); \(\theta\) labels the candidate prescription and is not an index.
Then
\begin{equation}
P_\theta^{\disc}(x^+|x)
=
\frac{
|\det(I-C_\theta)|
}{
(2\pi\Dt)^{d/2}|\det\sigma(\bar x_\theta)|
}
\exp\!\left[
-\frac{\Dt}{2}\eta_{\theta k}\eta_{\theta k}
\right].
\label{eq:exact-discrete-kernel}
\end{equation}

Equation~\eqref{eq:exact-discrete-kernel} is exact on a locally invertible branch of the chosen finite-step \(\theta\)-scheme whenever the map from \(\eta\) to \(x^+\) is locally invertible and \(\det\sigma\neq0\).  It is therefore the object used in the likelihood calculations.  By contrast, Eq.~\eqref{eq:Ltheta} is the contracted short-time action used to expose the local Jacobian structure and to define the continuum path functional.  Transition probabilities for multiplicative-noise processes require careful treatment of discretization and Jacobian factors \cite{LauLubensky2007,MorenoBarci2019,CugliandoloLecomte2017}.

In the numerical likelihood evaluation, nonfinite candidates or locally noninvertible branches are rejected.  Although Eq.~\eqref{eq:exact-discrete-kernel} contains \(|\det(I-C_\theta)|\), the stable parameter regimes used below have \(\det(I-C_\theta)>0\) along the accepted trajectories.  The implementation therefore assigns \(-\infty\) log likelihood to nonpositive or nonfinite determinant values rather than repairing them by hand.

\section{Matrix-Jacobian visibility}
\label{sec:visibility}

Recall the scalar \(\Jsig\) defined in Eq.~\eqref{eq:Jdef} for a specified noise-amplitude representation \(\sigma\).
The last term in Eq.~\eqref{eq:Ltheta} is \(\theta^2\Jsig/2\).  The two contractions in Eq.~\eqref{eq:Jdef} have different origins: the full contraction \(\partial_j\sigma_{ik}\partial_i\sigma_{jk}\) comes from \(\tr C^2\), while the divergence-square contraction comes from completing the square.

The scalar \(\Jsig\) vanishes in several common cases.  In one dimension both contractions reduce to \((\sigma')^2\).  For scalar-isotropic noise, \(\sigma_{ij}(x)=s(x)\delta_{ij}\), one finds \(\partial_j\sigma_{ik}\partial_i\sigma_{jk}=\partial_i s\,\partial_i s=(\partial_i\sigma_{ik})(\partial_l\sigma_{lk})\).  For strictly diagonal noise, \(\sigma_{ij}(x)=s_i(x)\delta_{ij}\), both contractions reduce to \(\sum_i(\partial_i s_i)^2\).  Thus \(\Jsig\) is invisible in many examples used to build intuition.

It can be nonzero for mixed matrix-valued noise.  A simple sufficient condition is the presence of an off-diagonal, state-dependent entry whose gradient couples different state rows through the same noise channel, so that \(\partial_j\sigma_{ik}\partial_i\sigma_{jk}\) contains a cross-row product not cancelled by \((\partial_i\sigma_{ik})(\partial_l\sigma_{lk})\).  Model A below realizes the minimal two-dimensional version: \(\sigma_{11}=1+by\) and \(\sigma_{21}=cx\) give \(\Jsig=2bc\).

One way to interpret this structure is as a local noise-amplitude frame whose columns are the directions in state space driven by each independent noise source.  In the Model A form, the first noise channel drives both coordinates with state-dependent projections, and the gradients of those projections couple the two state rows.  Similar local frames can appear when anisotropic diffusion or mobility tensors are projected onto coarse variables, when constrained dynamics is written in non-principal coordinates, or when fast variables are eliminated and leave state-dependent correlated fluctuations.  The covariance \(K\) then describes the local second moments, but the specified amplitude frame \(\sigma\) also carries derivative information that can enter the short-time path measure through \(\Jsig\).

\section{Inference of the stochastic prescription}
\label{sec:inference}

The generator equivalence in Sec.~\ref{sec:generator} shows that a density evolution cannot by itself separate the prescription parameter from a compensating change in drift.  We therefore treat \(\theta\) as a path-level parameter within a fixed finite-step model: the drift \(F\), the noise amplitude \(\sigma\), the time step \(\Dt\), and the interpolation rule are specified, and different candidate values of \(\theta\) assign different likelihoods to the same observed increments.

Given a trajectory \(x_0,\ldots,x_N\), define the discrete path log likelihood
\begin{equation*}
\mathcal L_N^{\disc}(\theta)
=
\sum_{n=0}^{N-1}
\log P_\theta^{\disc}(x_{n+1}|x_n;\Dt),
\end{equation*}
where the one-step density is the finite-step kernel in Eq.~\eqref{eq:exact-discrete-kernel}, not the contracted short-time action in Eq.~\eqref{eq:Ltheta}.  The inferred stochastic prescription is
\begin{equation}
\hat\theta
=
\arg\max_{\theta\in[0,1]}
\mathcal L_N^{\disc}(\theta).
\label{eq:theta-hat}
\end{equation}
At the population level this maximization is a Kullback--Leibler projection of the observed transition law onto the model family \(P_\theta^{\disc}\).  If the specified model family contains the generating finite-step process, the population maximizer recovers the generating prescription.  If a term in the likelihood is deliberately removed, the maximizer should instead be interpreted as the best misspecified member of the reduced family.

The numerical test below uses this latter idea as a diagnostic rather than as a claim about fitting unrelated physical models.  We evaluate two likelihoods on exactly the same trajectories: the full finite-step likelihood and a reduced-determinant control in which only the off-diagonal matrix contribution to \(\det(I-C_\theta)\) is suppressed.  Holding the Gaussian increment term, drift, diffusion matrix, and interpolation point fixed makes the resulting shift in \(\hat\theta\) a paired measure of the matrix-Jacobian contribution.

\subsection{Model A: paired inference shift}

We now test whether the matrix determinant contribution affects Eq.~\eqref{eq:theta-hat}.  Model A is intentionally minimal: it isolates the effect of removing only the off-diagonal determinant contribution while keeping the remaining likelihood structure fixed.  Consider \(F(x,y)=(-\kappa x,-\kappa y)\) and
\begin{equation}
\sigma(x,y)
=
\begin{pmatrix}
1+by & 0\\
cx & 1
\end{pmatrix}.
\label{eq:modelA}
\end{equation}
For this model \(a_k=\partial_i\sigma_{ik}=0\) and \(\Jsig=2bc\).
Writing \(H_\theta=I-C_\theta\) for the determinant matrix in Eq.~\eqref{eq:exact-discrete-kernel}, the exact full determinant reduces to
\begin{equation}
\det H_{\rm full}
=
(1+\theta\kappa\Dt)^2
-
\theta^2bc\,\eta_1^2\Dt^2.
\label{eq:modelA-full-det}
\end{equation}
As a clean control we remove only the off-diagonal matrix contribution and use
\begin{equation}
\det H_{\rm diag}
=
(1+\theta\kappa\Dt)^2.
\label{eq:modelA-diag-det}
\end{equation}
Thus the two likelihoods are identical when \(bc=0\), and differ only by the off-diagonal matrix determinant term when \(bc\neq0\).  The subscript ``diag'' labels this reduced determinant, not a diagonal diffusion matrix.  Because the reduced determinant likelihood intentionally removes a term present in the generating finite-step model, its maximizer is a misspecified KL projection; the quantity reported below is therefore a paired estimator shift, not a classical estimator bias.

\begin{figure}[t]
\centering
\includegraphics[width=0.92\linewidth]{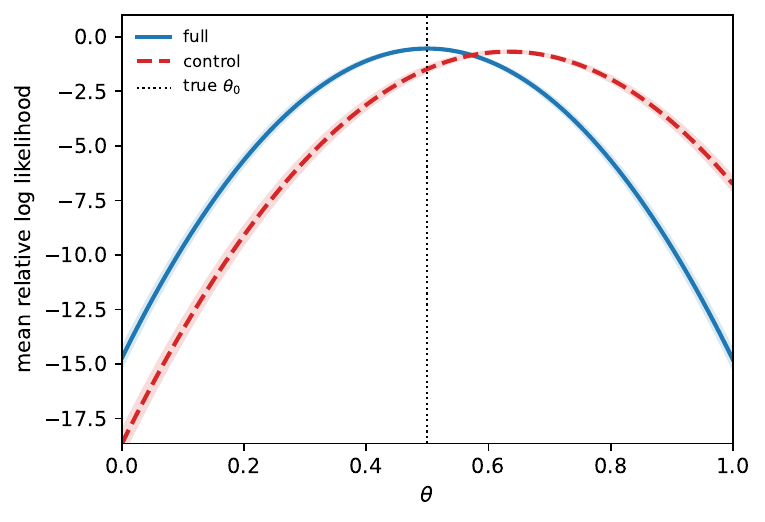}
\caption{Representative likelihood curves for Model A.  The full determinant likelihood and the reduced-determinant control are evaluated on the same trajectory.  The omitted off-diagonal determinant contribution shifts the maximum of the likelihood in \(\theta\).}
\label{fig:modelA-likelihood}
\end{figure}

Synthetic data are generated from the exact implicit \(\theta_0\)-scheme for Eq.~\eqref{eq:modelA}.  For this model the implicit update is linear in the increment and can be solved exactly at each time step, as described in Appendix~\ref{sec:S-modelA}.  We then fit \(\theta\) using both Eq.~\eqref{eq:modelA-full-det} and Eq.~\eqref{eq:modelA-diag-det}.  A representative likelihood pair is shown in Fig.~\ref{fig:modelA-likelihood}.

\begin{figure}[t]
\centering
\includegraphics[width=0.92\linewidth]{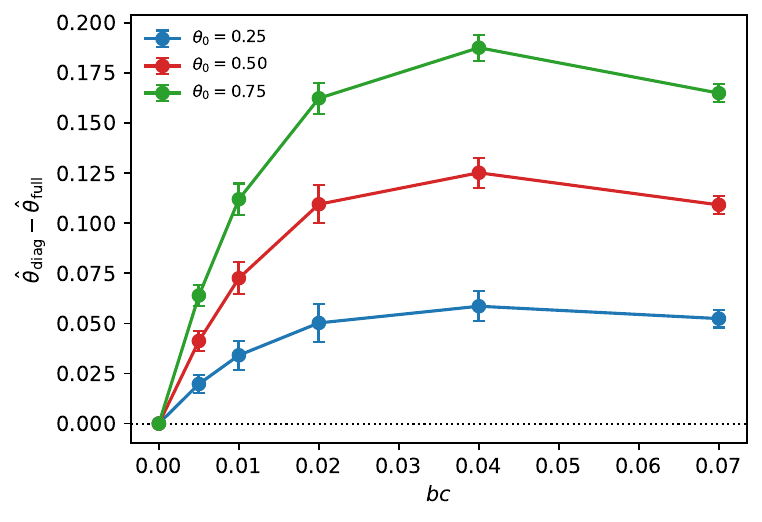}
\caption{Robustness of the paired inference shift in Model A under changes of the generating prescription.  The plotted quantity is the paired estimator shift \(\hat\theta_{\rm diag}-\hat\theta_{\rm full}\) for \(\theta_0=0.25,0.50,0.75\).  The shift vanishes at \(bc=0\) and remains visible for nonzero \(bc\), showing that the effect is not tied to the Stratonovich value \(\theta_0=1/2\).  Error bars denote SEM over independent seeds.}
\label{fig:modelA-theta0-sweep}
\end{figure}

The production paired-shift and robustness sweep is summarized in Fig.~\ref{fig:modelA-theta0-sweep} and Appendix~\ref{sec:S-modelA}.  The absolute maximum-likelihood estimate of \(\theta\) in Table~\ref{tab:S-modelA-summary} is not, by itself, the main evidence: individual finite records can show substantial sampling displacement.  To check that this displacement is not a simulator--likelihood mismatch, we performed a separate self-consistency validation of the full finite-step likelihood.  Increasing the trajectory length at fixed \(\Dt=10^{-3}\) gives \(\langle\hat\theta_{\full}\rangle=0.488\pm0.010\) at \(N=10^6\), while decreasing the step size to \(\Dt=5\times10^{-4}\) at comparable physical time gives \(\langle\hat\theta_{\full}\rangle=0.532\pm0.016\).  The population log likelihood, averaged over 101 independent trajectories, is maximized at \(\theta=0.500\).  A direct noise-reconstruction test also reproduces the driving Gaussian increments to numerical precision, with maximum absolute error \(1.4\times10^{-13}\).  Thus the absolute displacement of individual finite records reflects finite-sample variability rather than a mismatch between the simulator and the likelihood.

Figure~\ref{fig:modelA-theta0-sweep} gives the main Model A paired-shift diagnostic by repeating the comparison for \(\theta_0=0.25,0.50,0.75\).  The paired shift remains zero at \(bc=0\) and positive for the tested nonzero \(bc\) values across the three generating prescriptions.  Thus the observed effect is not an artifact of choosing the Stratonovich value \(\theta_0=1/2\).  For representative nonzero \(bc=0.04\), the paired shift is many SEMs away from zero; the corresponding standardized shifts are reported in Appendix~\ref{sec:S-modelA}.

The operational diagnostic is therefore the paired comparison on the same trajectory: the full determinant likelihood is compared with a reduced-determinant control that removes only the off-diagonal matrix determinant contribution while leaving the interpolation point, drift, diffusion matrix, and Gaussian increment term unchanged.  This paired design isolates the determinant contribution more cleanly than comparing unrelated fitted models.  The paired shift vanishes at \(bc=0\), where \(\Jsig=0\), and becomes positive for all tested \(bc>0\).  The shift is not claimed to be linear or monotone in \(bc\).  At larger \(bc\), the determinant contribution no longer acts as a small linear perturbation of the quadratic part of the likelihood; it distorts the finite-step likelihood surface nonlinearly, so the paired estimator shift should not be expected to grow monotonically with \(bc\).

\section{Optimal paths}
\label{sec:paths}

The likelihood test in Sec.~\ref{sec:inference} probes the finite-step determinant contribution through fitted prescriptions.  A complementary question is whether the same short-time Jacobian term changes the geometry of a most likely path when the stochastic model is treated as fixed.  This is a natural Onsager--Machlup application \cite{OnsagerMachlup1953,LiDuanLiu2021}: instead of asking which \(\theta\) best fits a trajectory, we ask whether the corrected action selects a different transition path between fixed endpoints.

For this variational problem we use the contracted short-time action in Eq.~\eqref{eq:Ltheta}, because the object of interest is the local continuum path functional rather than a finite-step likelihood over observed data.  Write Eq.~\eqref{eq:Ltheta} as
\begin{equation*}
\mathcal L_\theta(x,\dot x)
=
\frac12(\dot x_i-U_i)K^{-1}_{ij}(\dot x_j-U_j)
+
\Phi_\theta(x),
\end{equation*}
where
\begin{equation*}
U_i
=
F_i-\theta\sigma_{ik}\partial_m\sigma_{mk},
\qquad
\Phi_\theta
=
\theta\partial_iF_i+\frac{\theta^2}{2}\Jsig.
\end{equation*}
The canonical momentum is
\begin{equation*}
p_i=K^{-1}_{ij}(\dot x_j-U_j),
\qquad
\dot x_i=U_i+K_{ij}p_j.
\end{equation*}
To avoid confusion with the determinant matrix \(H_\theta=I-C_\theta\), we denote the path Hamiltonian by
\begin{equation*}
\mathcal H_\theta
=
p_iU_i+\frac12p_iK_{ij}p_j-\Phi_\theta.
\end{equation*}
Hamilton's equations give
\begin{equation}\label{eq:hamilton-p}
\dot p_i
=
-p_j\partial_iU_j
-\frac12p_jp_k\partial_iK_{jk}
+
\partial_i\Phi_\theta.
\end{equation}
Thus a state-dependent \(\Jsig\) contributes the force-like term \((\theta^2/2)\partial_i\Jsig\) to the optimal-path equation.

\subsection{Model B: path deformation}

To isolate this effect, Model B uses a deliberately simple variational setting.  The question is whether a state-dependent matrix-Jacobian contribution can deform a stable optimized path under a paired reduced-action comparison.  The design parallels Model A: the full and reduced problems keep the same drift, diffusion matrix, endpoints, time horizon, and discretized kinetic term, while differing only by the matrix-Jacobian scalar in the local action.  Consider
\begin{equation*}
\sigma(x,y)
=
\begin{pmatrix}
1+by & 0\\
cx^2 & 1
\end{pmatrix}.
\end{equation*}
For this model \(a_k=0\) and \(\Jsig=4bcx\).
Therefore the full action differs from the reduced action by
\begin{equation}
\Delta\mathcal L
=
\frac{\theta^2}{2}\Jsig
=
2\theta^2bcx.
\label{eq:modelB-deltaL}
\end{equation}
For \(bc>0\), Eq.~\eqref{eq:modelB-deltaL} lowers the action on the negative-\(x\) side and raises it on the positive-\(x\) side.  With endpoints on the \(y\)-axis, this gives a direct sign prediction: the full optimized path should bow into negative \(x\), whereas the reduced action, which lacks this linear-in-\(x\) term, should remain essentially vertical.  The numerical implementation is described in Appendix~\ref{sec:S-modelB}.

\begin{figure}[t]
\centering
\includegraphics[width=0.92\linewidth]{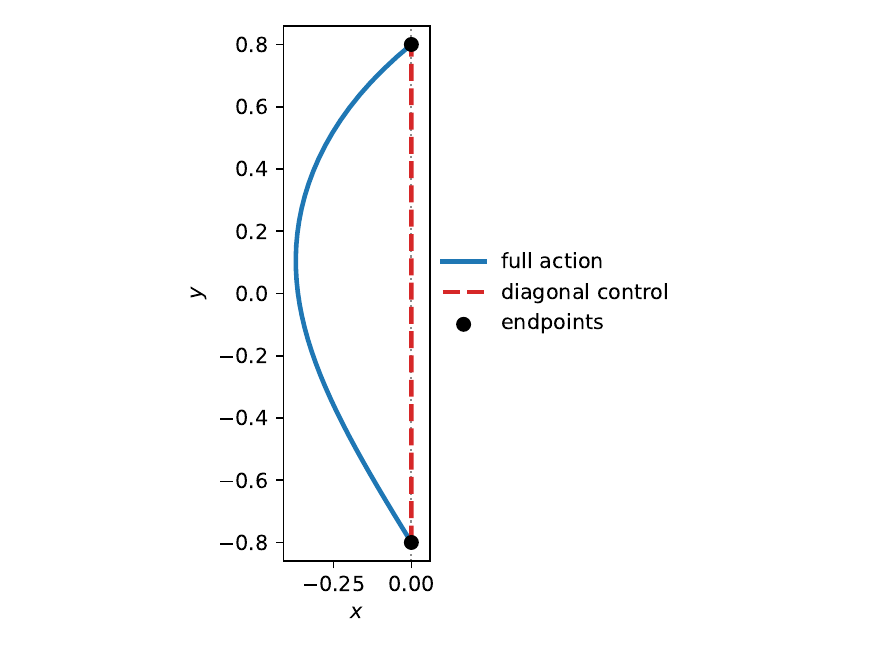}
\caption{Most likely transition paths for Model B.  The full action contains the state-dependent matrix-Jacobian term \(\Jsig=4bcx\), while the reduced-action control omits it.  The full optimal path bows into negative \(x\), as predicted by Eq.~\eqref{eq:modelB-deltaL}.}
\label{fig:modelB-path}
\end{figure}

Figure~\ref{fig:modelB-path} shows the resulting deformation.  For \(N=40\), the full action evaluated on the full path is \(-1.204223\), whereas the same full action evaluated on the path optimized under the reduced action is \(-1.053531\).  Conversely, the reduced action is minimized by its own optimized path: \(S_{\rm diag}=-1.053530\) on that path compared with \(-0.908849\) on the full path.  These action values need not be positive because the Onsager--Machlup functional contains divergence and Jacobian contributions; only action differences between paths evaluated with the same functional are used here.  The full path reaches \(x_{\min}=-0.371240\), while the path optimized under the reduced action remains on the axis to within \(10^{-6}\).  The \(N=60\) consistency run gives the same values to the shown precision; see Appendix~\ref{sec:S-modelB}.  These cross-action checks show that each reported path is favored by its own action.  To test whether the bowed solution is an initialization artifact, we repeated the \(N=40\) optimization from a cold straight-line initialization, from positive- and negative-bow perturbations, from a small sinusoidal \(x\) perturbation, and from five small random perturbations with fixed seeds.  The full action converged in all cases to the same bowed path within numerical tolerance; in particular, the cold-start full-action path differed from the negative-bow full-action solution by \(\sim 5\times10^{-6}\) in path norm.  The reduced action relaxed to the nearly vertical path even when initialized from the bowed manifold, with the bowed reduced-action initialization returning to the straight reduced-action solution within \(7.7\times10^{-7}\) in path norm.  We therefore interpret the deformation in Fig.~\ref{fig:modelB-path} as a stable feature of the full variational landscape rather than as a consequence of biased initialization.  These tests support the stability of the observed optimizer basin but should not be interpreted as an exhaustive proof of global uniqueness.

\section{Discussion}
\label{sec:discussion}

The central message is that a matrix part of the path-measure Jacobian can have operational consequences when it is tested under paired controls.  The effect is not inferred by comparing unrelated stochastic models, but by holding the drift, diffusion matrix, interpolation rule, Gaussian increment term, and numerical setting fixed while removing only the matrix-Jacobian contribution.  Because \(\Jsig\) vanishes in one-dimensional, scalar-isotropic, and diagonal-noise examples, this possibility is easy to miss if one extrapolates from scalar intuition.

The results also clarify what can and cannot be inferred.  The stochastic prescription is not separately identifiable from the Fokker--Planck generator without additional constraints.  Path likelihoods provide such constraints only after a model class is specified.  The inferred \(\theta\) should therefore be interpreted as a model-dependent path-level estimate, or more generally as a KL projection of observed transitions onto a \(\theta\)-discretized family.  In the present numerics, removing the matrix determinant contribution within a fixed path-likelihood model changes the fitted prescription or the stable optimized path while the rest of the model is held fixed.  For Model A, separate validation confirms that the full finite-step likelihood is self-consistent and recovers the generating prescription in aggregate, while suppressing the matrix determinant contribution produces a stable paired displacement of about \(0.12\)--\(0.15\) in the fitted convention parameter in the main production regime.  For Model B, the reported multi-start tests support stability of the observed optimizer basin, but they are not an exhaustive proof of global uniqueness.

These examples should be read with their scope in mind.  Many important multidimensional systems reduce effectively to scalar, isotropic, or diagonal structures where \(\Jsig\) vanishes, and even when \(\Jsig\neq0\) its numerical effect need not be large.  The present contribution is to identify a structurally distinct remainder and show that it can produce measurable paired shifts or path deformations in these settings.

A natural next step is to apply the same path-likelihood framework to microscopic regularizations in which an effective stochastic prescription is expected to vary with hidden time scales or feedback delays \cite{SanchoSanMiguelDurr1982,KupfermanPavliotisStuart2004,Pesce2013}.  Such a study would define a resolution-dependent \(\theta_{\eff}(\Dt)\), but we leave that spectroscopy problem for future work.
Other directions include broader model classes, rare-event sampling comparisons, coarse-grained physical systems, higher-dimensional inference, and manifold or constrained stochastic dynamics.

\appendix

\section{Short-time action and determinant expansion}
\label{sec:S-action}

This appendix gives the short-time expansion leading to Eq.~\eqref{eq:Ltheta}.  The point of keeping the calculation here is to make explicit which terms are discarded by ordinary scalar intuition and which terms survive because multiplicative-noise increments scale as \(\Delta x=O(\Dt^{1/2})\).  In particular, the derivation tracks the two contractions whose difference defines \(\Jsig\): one generated by \(\tr C^2\), and one generated by completing the square in the Gaussian noise variable.

Starting from the finite-step update Eq.~\eqref{eq:white-update}, define
\begin{equation*}
J_{il}=\frac{\partial x_i^+}{\partial \eta_l}.
\end{equation*}
Differentiating the implicit relation gives
\begin{equation*}
J_{il}
=
\theta\Dt
\left[
\partial_jF_i+\partial_j\sigma_{ik}\eta_k
\right]J_{jl}
+\sigma_{il}\Dt,
\end{equation*}
with all coefficients evaluated at \(\bar x\).  With
\begin{equation*}
C_{ij}=\theta\Dt\left[\partial_jF_i+\partial_j\sigma_{ik}\eta_k\right],
\end{equation*}
one has, up to a transpose convention that does not affect determinants,
\begin{equation*}
J=(I-C)^{-1}\sigma\Dt.
\end{equation*}
Thus the change of variables from \(\eta\) to \(x^+\) contributes
\begin{equation*}
\frac{1}{|\det J|}
=
\frac{|\det(I-C)|}{|\det\sigma|\Dt^d}.
\end{equation*}
For the locally invertible branch considered in the short-time expansion we take the logarithm of \(\det(I-C)\); the absolute value is retained in the finite-step kernel Eq.~\eqref{eq:exact-discrete-kernel}.

The scaling of the two pieces of \(C\) is different.  The drift-gradient part \(\theta\Dt\partial_jF_i\) is \(O(\Dt)\), whereas the multiplicative-noise part \(\theta\Dt\partial_j\sigma_{ik}\eta_k\) is \(O(\Dt^{1/2})\), because \(\eta_k=O(\Dt^{-1/2})\).  Consequently the linear trace of \(C\) and the square of the multiplicative-noise part of \(C\) both contribute to the exponent at order \(\Dt\).  Higher powers of \(C\) are \(O(\Dt^{3/2})\) or smaller in the short-time action.

Using
\begin{equation*}
\log\det(I-C)=-\tr C-\frac12\tr C^2+O(\Dt^{3/2}),
\end{equation*}
the first trace is
\begin{equation*}
\tr C
=
\theta\partial_iF_i\Dt
+
\theta\partial_i\sigma_{ik}\eta_k\Dt.
\end{equation*}
It is convenient to define the noise-channel divergence
\begin{equation*}
a_k=\partial_i\sigma_{ik}.
\end{equation*}
The second trace is
\begin{align*}
\tr C^2
&=C_{ij}C_{ji}
\\
&=
\theta^2\Dt^2
\left[
\partial_jF_i+\partial_j\sigma_{ik}\eta_k
\right]
\left[
\partial_iF_j+\partial_i\sigma_{jl}\eta_l
\right].
\end{align*}
The two terms containing one or two drift gradients are respectively \(O(\Dt^{3/2})\) and \(O(\Dt^2)\) in the exponent and do not contribute to the order kept here.  The only surviving part is therefore
\begin{equation*}
\tr C^2
=
\theta^2
\partial_j\sigma_{ik}\partial_i\sigma_{jl}
\eta_k\eta_l\Dt^2
+O(\Dt^{3/2}).
\end{equation*}
For typical short-time increments, the L\'evy contraction gives
\begin{equation*}
\eta_k\eta_l\Dt^2
=
\Delta W_k\Delta W_l
\longrightarrow
\delta_{kl}\Dt,
\end{equation*}
inside the order-\(\Dt\) action.  This yields
\begin{equation*}
\tr C^2
\longrightarrow
\theta^2
\partial_j\sigma_{ik}\partial_i\sigma_{jk}\Dt.
\end{equation*}
Defining
\begin{equation*}
\Msig=\partial_j\sigma_{ik}\partial_i\sigma_{jk},
\end{equation*}
the determinant contribution becomes
\begin{multline*}
\log\det(I-C)
=
-\theta\partial_iF_i\Dt
-\theta a_k\eta_k\Dt \\
-\frac{\theta^2}{2}\Msig\Dt
+O(\Dt^{3/2}).
\end{multline*}

Now write
\begin{equation*}
\eta_k=\sigma^{-1}_{ki}(v_i-F_i),
\qquad
v_i=\frac{x_i^+-x_i}{\Dt},
\end{equation*}
again with coefficients evaluated at the interpolation point.  Combining the Gaussian factor \(-\Dt\eta_k\eta_k/2\) with the linear determinant term gives
\begin{equation*}
-\frac12\eta_k\eta_k-\theta a_k\eta_k
=
-\frac12(\eta_k+\theta a_k)(\eta_k+\theta a_k)
+
\frac{\theta^2}{2}a_ka_k.
\end{equation*}
The shifted noise variable corresponds to the shifted velocity combination
\begin{equation*}
\eta_k+\theta a_k
=
\sigma^{-1}_{ki}
\left(v_i-F_i+\theta\sigma_{ik}\partial_m\sigma_{mk}\right).
\end{equation*}
Substituting this square and the remaining order-\(\Dt\) determinant terms into the exponent gives
\begin{widetext}
\begin{equation*}
\mathcal L_\theta(x,v)
=
\frac12
\left(
 v_i-F_i+\theta\sigma_{ik}\partial_m\sigma_{mk}
\right)
\sigma^{-1}_{li}\sigma^{-1}_{lj}
\left(
 v_j-F_j+\theta\sigma_{jn}\partial_p\sigma_{pn}
\right)
+
\theta\partial_iF_i
+
\frac{\theta^2}{2}
\left[
\partial_j\sigma_{ik}\partial_i\sigma_{jk}
-
(\partial_i\sigma_{ik})(\partial_l\sigma_{lk})
\right].
\end{equation*}
\end{widetext}
The first contraction in the last bracket comes from \(\tr C^2\), while the second comes from completing the square in the Gaussian noise variable.  These two contractions coincide in one-dimensional, scalar-isotropic, and strictly diagonal examples, but they are not equal for a generic matrix-valued noise amplitude.  Thus
\begin{equation*}
\Jsig=
\partial_j\sigma_{ik}\partial_i\sigma_{jk}
-
(\partial_i\sigma_{ik})(\partial_l\sigma_{lk})
\end{equation*}
is the part of the short-time Jacobian contribution that is hidden in those simpler cases.

\section{Model A: simulation and likelihood fitting}
\label{sec:S-modelA}

Model A has \(F(x,y)=(-\kappa x,-\kappa y)\) and
\begin{equation*}
\sigma(x,y)=
\begin{pmatrix}
1+by & 0\\
cx & 1
\end{pmatrix}.
\end{equation*}
The inverse and determinant are
\begin{equation*}
\sigma^{-1}=
\begin{pmatrix}
(1+by)^{-1} & 0\\
-cx(1+by)^{-1} & 1
\end{pmatrix},
\qquad
\det\sigma=1+by.
\end{equation*}
The nonzero derivatives are \(\partial_y\sigma_{11}=b\) and \(\partial_x\sigma_{21}=c\).  Hence \(a_k=\partial_i\sigma_{ik}=0\) and \(\Msig=\Jsig=2bc\).

\paragraph*{Exact implicit simulator.}

Here ``exact'' refers to the algebraic solution of the chosen finite-step implicit update, not to an exact transition density of the continuous-time diffusion.

For the algebraic finite-step solve at true prescription \(\theta_0\), write the increments as \(\Delta x=dx\) and \(\Delta y=dy\).  The implicit scheme is
\begin{align*}
dx
&=
-\kappa(x+\theta_0 dx)\Dt
+[1+b(y+\theta_0dy)]\Delta W_1,
\\
dy
&=
-\kappa(y+\theta_0dy)\Dt
+c(x+\theta_0dx)\Delta W_1+\Delta W_2.
\end{align*}
This gives the linear system
\begin{equation*}
M
\begin{pmatrix}
dx\\ dy
\end{pmatrix}
=
\begin{pmatrix}
r_x\\ r_y
\end{pmatrix},
\end{equation*}
where
\begin{equation*}
M=
\begin{pmatrix}
\alpha & -b\theta_0\Delta W_1\\
-c\theta_0\Delta W_1 & \alpha
\end{pmatrix},
\qquad
\alpha=1+\kappa\theta_0\Dt,
\end{equation*}
\begin{align*}
r_x &= -\kappa x\Dt+(1+by)\Delta W_1, \\
r_y &= -\kappa y\Dt+cx\Delta W_1+\Delta W_2.
\end{align*}
The determinant of the implicit solve is
\begin{equation*}
D_M\equiv\det M
=
\alpha^2-bc\theta_0^2(\Delta W_1)^2 .
\end{equation*}
Hence the exact finite-step increment used by the simulator is
\begin{equation}
dx
=
\frac{\alpha r_x+b\,\theta_0 r_y \Delta W_1}{D_M},
\qquad
dy
=
\frac{\alpha r_y + c\,\theta_0 r_x \Delta W_1}{D_M}.
\label{eq:S-modelA-exact-increment}
\end{equation}
Equation~\eqref{eq:S-modelA-exact-increment} is the form used in our simulations.

\paragraph*{Full and reduced-control determinants.}

For candidate \(\theta\), writing \(H_\theta=I-C_\theta\), the Model A full and reduced-control determinants are given in Eqs.~\eqref{eq:modelA-full-det} and \eqref{eq:modelA-diag-det}.  The reduced-determinant control keeps the exact diagonal part of the determinant but removes the off-diagonal matrix contribution.
This control differs from the full likelihood only through the genuine off-diagonal matrix determinant term.  The label \(\diag\) is retained only as a compact notation for this reduced determinant; it does not mean that the diffusion matrix is diagonal.  When \(bc=0\), the two likelihoods are identical.

\paragraph*{Simulation parameters.}

The simulation parameters for Model A are listed in Table~\ref{tab:S-modelA-params}. For each trajectory, an initial transient of \(20000\) steps was discarded before likelihood evaluation so that the reported statistics reflect the stationary regime rather than initialization effects.

\begin{table}[h]
\caption{Model A simulation parameters.}
\label{tab:S-modelA-params}
\begin{ruledtabular}
\begin{tabular}{ll}
Parameter & Value \\
\hline
\(\kappa\) & \(2.0\)\\
\(\theta_0\) & \(0.5\)\\
\(\Dt\) & \(10^{-3}\)\\
\(b\) & \(0.20\)\\
\(c\) & \(0,\,0.025,\,0.05,\,0.10,\,0.20,\,0.35\)\\
\(N_{\rm total}\) & \(300000\)\\
Initial transient removed & \(20000\)\\
\(\theta\)-grid & \(301\) points in \([0,1]\)\\
\end{tabular}
\end{ruledtabular}
\end{table}

\paragraph*{Inference summary.}

The paired inference shift is summarized in Table~\ref{tab:S-modelA-summary}.  The paired estimator shift is defined as \(\hat\theta_{\diag}-\hat\theta_{\full}\).
The absolute value of \(\hat\theta_{\full}\) in this finite set of production trajectories is not used as the primary diagnostic because individual finite records have appreciable sampling displacement.  The production runs in Table~\ref{tab:S-modelA-summary} were designed for paired sensitivity across \(bc\), not for high-precision absolute recovery.  A separate validation run confirms that the full finite-step likelihood recovers \(\theta_0\) in aggregate: the population likelihood is maximized at \(\theta=0.500\), the largest-\(N\) scaling run gives \(\langle\hat\theta_{\full}\rangle=0.488\pm0.010\), the smallest-\(\Dt\) run gives \(\langle\hat\theta_{\full}\rangle=0.532\pm0.016\), and direct reconstruction of the driving noise gives a maximum absolute error of \(1.4\times10^{-13}\).  The operational diagnostic in Table~\ref{tab:S-modelA-summary} remains the paired shift on the same trajectory.

\begin{table}[h]
\caption{Model A inference summary.  Each row uses ten independent seeds.  The table reports the paired shift between the reduced-determinant control and the full determinant likelihood.}
\label{tab:S-modelA-summary}
\begin{ruledtabular}
\begin{tabular}{cccccc}
\(bc\) & \(n\) & \(\langle\hat\theta_{\full}\rangle\) &
\(\langle\hat\theta_{\diag}\rangle\) &
shift mean & shift SEM\\
\hline
0.000 & 10 & 0.4542 & 0.4542 & \(-4.37\times10^{-11}\) & \(3.11\times10^{-10}\)\\
0.005 & 10 & 0.4551 & 0.4963 & 0.0412 & 0.0051\\
0.010 & 10 & 0.4568 & 0.5294 & 0.0726 & 0.0080\\
0.020 & 10 & 0.4611 & 0.5706 & 0.1095 & 0.0095\\
0.040 & 10 & 0.4691 & 0.5943 & 0.1252 & 0.0075\\
0.070 & 10 & 0.4771 & 0.5863 & 0.1092 & 0.0044\\
\end{tabular}
\end{ruledtabular}
\end{table}

\begin{table}[h]
\caption{Model A robustness under changes of the generating prescription.  The table reports the paired estimator shift for the representative value \(bc=0.04\).}
\label{tab:S-modelA-theta0-sweep}
\begin{ruledtabular}
\begin{tabular}{cccc}
\(\theta_0\) & \(n\) & shift mean \(\pm\) SEM & shift \(z\)\\
\hline
0.25 & 10 & \(0.0586\pm0.0075\) & 7.80\\
0.50 & 10 & \(0.1252\pm0.0075\) & 16.62\\
0.75 & 10 & \(0.1876\pm0.0066\) & 28.63\\
\end{tabular}
\end{ruledtabular}
\end{table}

\section{Model B: optimal-path calculation}
\label{sec:S-modelB}

Model B has \(F(x,y)=(-\kappa x,-\kappa y)\) and
\begin{equation*}
\sigma(x,y)=
\begin{pmatrix}
1+by & 0\\
cx^2 & 1
\end{pmatrix}.
\end{equation*}
Then \(a_k=0\) and \(\Jsig=4bcx\).
For Model B, Eq.~\eqref{eq:modelB-deltaL} gives \(\Delta\mathcal L=2\theta^2bcx\).
For \(bc>0\), this term lowers the action on the negative-\(x\) side.

\paragraph*{Discrete path action.}

For a path \(z_n=(x_n,y_n)\), \(n=0,\ldots,N\), with total time \(T\), we use 
$$ S=\sum_{n=0}^{N-1}\Delta t\,\mathcal L(z_{n+1/2},v_n), $$
where 
$$ z_{n+1/2}=\frac{z_n+z_{n+1}}{2},\quad v_n=\frac{z_{n+1}-z_n}{\Delta t},~\text{and} \quad \Delta t=\frac{T}{N}. $$
The full and reduced-control Lagrangians are
\begin{equation*}
\mathcal L_{\full}
=
\frac12(v-F)^TK^{-1}(v-F)+\theta\nabla\cdot F+\frac{\theta^2}{2}\Jsig,
\end{equation*}
and
\begin{equation*}
\mathcal L_{\diag}
=
\frac12(v-F)^TK^{-1}(v-F)+\theta\nabla\cdot F.
\end{equation*}
For Model B,
\begin{equation*}
\sigma^{-1}=
\begin{pmatrix}
(1+by)^{-1} & 0\\
-cx^2(1+by)^{-1} & 1
\end{pmatrix},
\qquad
K^{-1}=\sigma^{-T}\sigma^{-1}.
\end{equation*}

\paragraph*{Hamiltonian form.}

The Hamiltonian formulation is not used as the numerical optimizer in Fig.~\ref{fig:modelB-path}; the actual computation minimizes the discrete midpoint action below.  It is included here to identify the variational force produced by the matrix-Jacobian term.  With the shifted drift and scalar potential defined in Sec.~\ref{sec:paths},
\begin{equation*}
U_i=F_i-\theta\sigma_{ik}\partial_m\sigma_{mk},
\qquad
\Phi_\theta=\theta\partial_iF_i+\frac{\theta^2}{2}\Jsig,
\end{equation*}
the momentum conjugate to the path is
\begin{equation*}
p_i=K^{-1}_{ij}(\dot x_j-U_j),
\qquad
\dot x_i=U_i+K_{ij}p_j.
\end{equation*}
The corresponding path Hamiltonian is
\begin{equation*}
\mathcal H_\theta
=
p_iU_i+\frac12p_iK_{ij}p_j-\Phi_\theta,
\end{equation*}
which is a variational Hamiltonian and is distinct from the determinant matrix \(H_\theta=I-C_\theta\) used in the finite-step likelihood.  Taking the Hamilton equation for \(p_i\) gives the usual drift-gradient and metric-gradient terms, plus the potential-gradient contribution
\begin{equation*}
\partial_i\Phi_\theta
=
\theta\partial_i\partial_jF_j+\frac{\theta^2}{2}\partial_i\Jsig.
\end{equation*}
Thus the only difference between the full and reduced Model B path equations is the additional force-like term
\begin{equation*}
\frac{\theta^2}{2}\partial_i\Jsig.
\end{equation*}
For \(\Jsig=4bcx\), the potential-gradient contribution entering Hamilton's equation is
\begin{equation*}
\frac{\theta^2}{2}\nabla\Jsig=(2\theta^2bc,0).
\end{equation*}
Equivalently, the action decreases in the direction
\begin{equation*}
-\nabla\Phi_\theta=(-2\theta^2bc,0),
\end{equation*}
so for \(bc>0\) the variational minimum is biased toward negative \(x\).  This is the variational origin of the leftward bowing of the full optimized path.

\paragraph*{Simulation parameters and action checks.}

The endpoints are \(A=(0,-0.8)\) and \(B=(0,0.8)\).  The parameters are \(\theta=1\), \(\kappa=0.25\), \(b=0.25\), \(c=0.8\), and \(T=3\).
The optimization uses L-BFGS-B over interior path points with bounds \(x,y\in[-2,2]\).  For the optimization shown in Fig.~\ref{fig:modelB-path}, the full-action path was initialized using a small negative-\(x\) bow,
\(x_{\rm init}(t)\to x_{\rm init}(t)-0.15\sin(\pi t/T)\), while the reduced-control path was initialized as a straight line.
A large penalty is returned if \(1+by_{n+1/2}\le0.05\).

Here \(S_{\full}[z]\) and \(S_{\diag}[z]\) denote the full and reduced-control actions evaluated on path \(z\), while \(z_{\full}\) and \(z_{\diag}\) are the corresponding optimized paths.  The \(N=40\) action checks are shown in Table~\ref{tab:S-modelB-actions}.

\begin{table}[h]
\caption{Model B action checks for \(N=40\).}
\label{tab:S-modelB-actions}
\begin{ruledtabular}
\begin{tabular}{lcc}
Quantity & \(z_{\full}\) & \(z_{\diag}\)\\
\hline
\(S_{\full}[z]\) & \(-1.204223\) & \(-1.053531\)\\
\(S_{\diag}[z]\) & \(-0.908849\) & \(-1.053530\)\\
\(\min x[z]\) & \(-0.371240\) & --\\
\(\max |x[z]|\) & -- & \(1\times10^{-6}\)\\
\end{tabular}
\end{ruledtabular}
\end{table}

The full action is lower on the bowed full path than on the path optimized under the reduced action, while the reduced action is lower on its own nearly vertical optimized path.  A consistency run with \(N=60\) gave \(S_{\full}[z_{\full}]=-1.204269\), \(S_{\full}[z_{\diag}]=-1.053525\), \(S_{\diag}[z_{\full}]=-0.908792\), and \(S_{\diag}[z_{\diag}]=-1.053524\), with \(\min x[z_{\full}]=-0.371326\) and \(\max |x[z_{\diag}]|=2\times10^{-6}\).  We subsequently performed a dedicated \(N=40\) cold-start and multi-start check using straight-line, positive-bow, negative-bow, sinusoidal, and fixed-seed random initializations.  The full action converged reproducibly to the same bowed solution, with the cold-start full-action path differing from the negative-bow full-action solution by \(\sim 5\times10^{-6}\) in path norm.  The reduced action converged to the nearly vertical solution for all tested starts, and the bowed reduced-action initialization relaxed back to the straight reduced-action solution within \(7.7\times10^{-7}\) in path norm.  These tests support the stability of the observed optimizer basin but should not be interpreted as an exhaustive proof of global uniqueness.

\section*{Data Availability}

The numerical scripts, generated data tables, and figure-generation assets supporting this study are available in Zenodo at DOI: \href{https://doi.org/10.5281/zenodo.20117517}{10.5281/zenodo.20117517}.

\bibliographystyle{apsrev4-2}
\bibliography{matrix_noise_stochastic_calculus}

\end{document}